\documentclass{article}

\usepackage{microtype}
\usepackage{graphicx}
\usepackage{subcaption}
\usepackage{booktabs} 

\usepackage{hyperref}

\usepackage[accepted]{icml2025}

\usepackage{amsmath}
\usepackage{amssymb}
\usepackage{mathtools}
\usepackage{amsthm}

\usepackage[capitalize,noabbrev]{cleveref}

\usepackage{tikz}
\usetikzlibrary{quantikz2}
\usepackage{blochsphere}

\usepackage{tablefootnote}

\theoremstyle{plain}

\theoremstyle{definition}

\theoremstyle{remark}

\usepackage[textsize=tiny]{todonotes}
\newcommand{\thickhline}{\noalign{\hrule height 1pt}}

\icmltitlerunning{Investigating the Lottery Ticket Hypothesis for Variational Quantum Circuits}

\begin{document}

\twocolumn[
\icmltitle{Investigating the Lottery Ticket Hypothesis \\for Variational Quantum Circuits}



\icmlsetsymbol{equal}{*}

\begin{icmlauthorlist}
\icmlauthor{Michael Kölle}{yyy}
\icmlauthor{Leonhard Klingert}{yyy}
\icmlauthor{Julian Schönberger}{yyy}
\icmlauthor{Philipp Altmann}{yyy} 
\icmlauthor{Tobias Rohe}{yyy}
\icmlauthor{Claudia Linnhoff-Popien}{yyy}
\end{icmlauthorlist}

\icmlaffiliation{yyy}{Institute of Informatics, LMU Munich, Munich, Germany}

\icmlcorrespondingauthor{Michael Kölle}{michael.koelle@ifi.lmu.de}

\icmlkeywords{Variational Quantum Circuits, Lottery Ticket Hypothesis, Barren Plateaus}
\vskip 0.3in
]



\printAffiliationsAndNotice{}  

\begin{abstract}
Quantum computing is an emerging field in computer science that has seen considerable progress in recent years, especially in machine learning. By harnessing the principles of quantum physics, it can surpass the limitations of classical algorithms. However, variational quantum circuits (VQCs), which rely on adjustable parameters, often face the barren plateau phenomenon, hindering optimization. The Lottery Ticket Hypothesis (LTH) is a recent concept in classical machine learning that has led to notable improvements in parameter efficiency for neural networks. It states that within a large network, a smaller, more efficient subnetwork, or “winning ticket,” can achieve comparable performance, potentially circumventing plateau challenges. In this work, we investigate whether this idea can apply to VQCs. We show that the weak LTH holds for VQCs, revealing winning tickets that retain just 26.0\% of the original parameters. For the strong LTH, where a pruning mask is learned without any training, we discovered a winning ticket in a binary VQC, achieving 100\% accuracy with only 45\% of the weights. These findings indicate that LTH may mitigate barren plateaus by reducing parameter counts while preserving performance, thus enhancing the efficiency of VQCs in quantum machine learning tasks.
\end{abstract}

\section{Introduction}
\label{sec:introduction}

Quantum computing (QC) has attracted significant attention due to its potential to surpass classical systems by leveraging superposition and entanglement, enabling solutions to problems once considered intractable \cite{nielsen2010quantum,NAP25196,sood2024archives}. VQCs, as quantum analogues of neural networks, have shown promise in tasks like classification, regression, and optimization \cite{cerezo2021variational,Du2022QuantumCircuitArchitectureSearch,Schuld2021vqc_as_ML_model,Qi2024VariationalQuantumAlgorithms}. Despite this promise, VQCs encounter barren plateaus (BP), where gradients become nearly zero, hindering parameter optimization \cite{mcclean2018barren,Cerezo2021CostFunctionBarrenPlateaus,grant2019initialization}.

The main objective of this work is to investigate whether the LTH, originally developed for classical neural networks \cite{frankle2019lottery}, can strengthen VQCs. The LTH posits that within an over-parameterized model, a smaller subnetwork, known as a “winning ticket,” can be trained in isolation to match or exceed the original model’s performance. One typically obtains a winning ticket by training a network, pruning low-magnitude weights, resetting the remaining weights to their initial values, and retraining \cite{frankle2019lottery,malach2020provinglotterytickethypothesis}. Pruning can be iterative or one-shot, and the strong LTH proposes that pruning alone can sometimes identify winning tickets without any training \cite{ramanujan2020whatshiddenrandomlyweighted,liu2024surveylotterytickethypothesis,cunha2022proving}.

Identifying winning tickets in VQCs may help mitigate BP issues by reducing circuit complexity while preserving performance. Smaller circuits could potentially exhibit less severe gradient decay, making optimization more feasible. This paper seeks to evaluate whether sparse subcircuits can indeed match the performance of full VQCs, thus offering a path to more efficient quantum ML.

In the remainder of this paper, \cref{sec:fundamentals_related_work} covers foundational ideas and related research in machine learning, pruning, the Lottery Ticket Hypothesis, quantum computing, and the barren plateau phenomenon. \cref{chap:Methods} details our methodology, including experiment design, pruning strategies, datasets, models, metrics, implementation details and hyperparameter optimization. \cref{sec:ResultsDiscussion} then presents and analyzes our experimental findings. Finally, \cref{sec:Conclusion} summarizes the paper and outlines avenues for future study.

\section{Related Work}
\label{sec:fundamentals_related_work}

Machine learning (ML) often relies on large neural networks (NNs) with millions of parameters to handle complex data, but these expansive models can incur high memory usage and computational overhead. Pruning has thus emerged as a way to remove parameters that exert minimal influence on a model’s predictions, retaining performance while reducing resource costs \cite{blalock2020stateneuralnetworkpruning}.

\subsection{Pruning in Classical Machine Learning}
Pruning typically involves producing a mask, a tensor of the same shape as the NN’s weights. Pruned parameters become zeros when multiplied by this mask. Magnitude pruning, which removes weights with small absolute values, is a widely used strategy. Pruning can be \emph{unstructured}—removing individual weights—or \emph{structured}—removing entire layers or channels, often leading to more substantial memory and runtime gains \cite{blalock2020stateneuralnetworkpruning}. In practice, pruning can proceed \emph{iteratively}, with repeated cycles of training and pruning until a target level of sparsity is reached, or via a \emph{one-shot} approach, where pruning happens once at a chosen fraction of weights, followed by optional fine-tuning \cite{frankle2019lottery,malach2020provinglotterytickethypothesis}.

\subsection{Lottery Ticket Hypothesis in Classical Machine Learning}
The LTH states that within a randomly initialized NN, there exists a sparse subnetwork (a “winning ticket”) which, when trained in isolation from its original initialization, can achieve performance comparable to or better than the full model \cite{frankle2019lottery}. In its \emph{weak} form, a network is trained and pruned, and the remaining weights are reset to their original values before a final retraining phase \cite{frankle2019lottery,malach2020provinglotterytickethypothesis}. The \emph{strong} variant posits that pruning alone can identify a strong-performing subnetwork without any initial training, as shown by algorithms that learn a mask directly, such as edge-popup \cite{ramanujan2020whatshiddenrandomlyweighted,malach2020provinglotterytickethypothesis}.

Studies validating the LTH in image classification, natural language processing, and other domains have revealed winning tickets that can retain only a small fraction of the original parameters while preserving or surpassing performance \cite{frankle2019lottery,gale2019state,liu2024surveylotterytickethypothesis}. The phenomenon has broad implications for model compression, transfer learning, and interpretability, since smaller subnetworks often train faster, adapt more efficiently to new tasks, and expose which parts of a network drive predictions \cite{cunha2022proving,ferbach2023generalframeworkprovingequivariant}.

\subsection{Barren Plateaus in Variational Quantum Circuits}
VQCs have emerged as a promising quantum counterpart to NNs, offering potential computational advantages in classification, regression, and optimization \cite{cerezo2021variational,Du2022QuantumCircuitArchitectureSearch,Qi2024VariationalQuantumAlgorithms}. Despite this promise, VQCs can suffer from BPs in their optimization landscapes, where gradients vanish and training fails to converge effectively \cite{mcclean2018barren}. Several factors, such as circuit depth, entanglement, cost function design, and parameter initialization, influence the severity of BPs \cite{Cerezo2021CostFunctionBarrenPlateaus,grant2019initialization,cunningham2024investigating}. Recommended strategies include careful parameter initialization, local cost functions, layer-wise training, gradient clipping, adaptive learning rates, quantum natural gradients, and regularization methods, all of which aim to reduce the likelihood of becoming trapped in flat regions of the parameter space \cite{cunningham2024investigating}.

\subsection{Pruning in Variational Quantum Circuits}
To mitigate BPs and improve efficiency, researchers have proposed pruning strategies for VQCs. For instance, Ma et al. \cite{ma2024continuous} present a continuous architecture search that prunes gates through a Structure Symmetric Pruning approach, which reduces both parameter counts and circuit depth. Similarly, Sim et al. \cite{sim2021adaptive} propose Parameter-Efficient Circuit Training (PECT), a technique that updates only a subset of parameters per iteration, thereby lowering runtime. Liu et al. \cite{kulshrestha2024qadapruneadaptiveparameterpruning} introduce Quantum Adaptive Pruning (QAdaPrune), which adaptively determines pruning thresholds to remove redundant parameters. These methods demonstrate that smaller VQCs can offer competitive performance, often improving trainability when BPs are present.

\subsection{Applying the LTH to Variational Quantum Circuits}
While pruning has found broad application in classical NNs, leveraging LTH-style sparsification in quantum circuits remains largely unexplored. Preliminary successes in pruning VQCs suggest that identifying small, high-performing subcircuits could mitigate BPs by reducing parameter redundancy and easing optimization demands \cite{sim2021adaptive,kulshrestha2024qadapruneadaptiveparameterpruning,ma2024continuous}. Strong and weak variants of the LTH could both be relevant: the former might uncover effective quantum subcircuits without extended training, while the latter would involve train-prune-retrain cycles typical of classical approaches \cite{frankle2019lottery,malach2020provinglotterytickethypothesis}. A successful transfer of the LTH to VQCs holds promise for more efficient quantum machine learning, addressing both computational resource constraints and the challenges of vanishing gradients.

\section{Methodology}\label{chap:Methods}
This section presents the methodology employed in this work to investigate the LTH in the context of VQCs. We discuss the hypothesis variants and the associated pruning techniques, describe the selected datasets and implemented models, and outline the experimental setup, including the programming languages and libraries used. Finally, we detail the considered hyperparameters and the applied evaluation metrics.

\subsection{Evaluating the Weak Lottery Ticket Hypothesis}\label{subsec:approach_weak}
Initially, we evaluated the weak LTH introduced by Frankle and Carbin (2019) \cite{frankle2019lottery}. We developed a training framework for both NNs and simulated VQCs, which also included a flexible pruning module that can operate in iterative or one-shot pruning modes. Both pruning approaches start with an unpruned model, and each pruning ratio spawns a new model. By setting a manual random seed before creating each model, we ensured identical initial randomized weights. We applied pruning before beginning each training iteration.

\subsubsection{Iterative Pruning}\label{subsec:approach_weak_iterative}
For iterative pruning, the model is repeatedly trained, pruned, and reset to evaluate performance at different remaining-weight levels \cite{frankle2019lottery,liu2024surveylotterytickethypothesis,malach2020provinglotterytickethypothesis}. \cref{alg:iterative} outlines our iterative pruning algorithm. First, we iterate over a list of random seeds to ensure reproducibility and mitigate seed-specific variations. In each iteration, we initialize a pruning mask that leaves all weights unpruned and set a variable to store the number of remaining weights. In the main loop, we set the random seed, create a new model, apply the pruning mask, and then train and evaluate the model. We subsequently update the pruning mask by pruning 20\% of the weights using magnitude pruning. Next, we measure the updated number of remaining weights. The loop terminates if either (1) the remaining weight count does not change due to insufficient weights for further pruning or (2) the number of remaining weights falls below a user-defined threshold that controls pruning depth.

\begin{algorithm}
    \caption{Weak Lottery Ticket Hypothesis - Iterative Pruning}
    \label{alg:iterative}
    \begin{algorithmic}[1]
        \STATE \textbf{Input:} list of $random\_seeds$, $RW_{threshold}$ for remaining weights
        \FOR{$random\_seed \in random\_seeds$}
            \STATE $PM \gets \text{initial pruning mask}$
            \STATE $RW_0 \gets -1$
            \STATE $i \gets 0$
            \REPEAT
                \STATE set $random\_seed$
                \STATE $M_i \gets \text{create new model}$
                \STATE apply PM on $M_i$
                \STATE train $M_i$
                \STATE evaluate $M_i$
                \STATE $i \gets i + 1$
                \STATE $PM \gets \text{update PM}$
                \STATE $RW_i \gets \text{measure remaining weights of } PM$
            \UNTIL{$RW_i \leq RW_{threshold} \text{ or } RW_i = RW_{i-1}$}
        \ENDFOR
    \end{algorithmic}
\end{algorithm}

\subsubsection{One-shot Pruning}\label{subsec:approach_weak_one_shot}
In the one-shot pruning approach, a set of pruning ratios is defined ahead of time \cite{frankle2019lottery,gale2019state,liu2024surveylotterytickethypothesis,malach2020provinglotterytickethypothesis}. The model is first trained in its unpruned state; then, separate copies are pruned according to each ratio and retrained from scratch, enabling direct performance comparisons across varying remaining-weight levels. \cref{alg:one_shot} shows our one-shot pruning algorithm. We begin by iterating over a list of random seeds. For each seed, we create, train, and evaluate an initial model $M_0$. Then, for each pruning rate, we set the same random seed, create a new model, prune it according to $M_0$'s weights, and retrain and evaluate the pruned model.

\begin{algorithm}
    \caption{Weak Lottery Ticket Hypothesis - One-shot Pruning}
    \label{alg:one_shot}
    \begin{algorithmic}[1]
        \STATE \textbf{Input:} list of $random\_seeds$, list of $pruning\_ratios$
        \FOR{$random\_seed \in random\_seeds$}
            \STATE set $random\_seed$
            \STATE $M_0 \gets \text{create initial model}$
            \STATE train $M_0$
            \STATE evaluate $M_0$
            \STATE $i \gets 0$
            \FOR{$pruning\_ratio \in pruning\_ratios$}
                \STATE $i \gets i + 1$
                \STATE $PM_i \gets \text{prune } M_0 \text{ by } pruning\_ratio\%$
                \STATE set $random\_seed$
                \STATE $M_i \gets \text{create new model}$
                \STATE apply $PM_i$ on $M_i$
                \STATE train $M_i$
                \STATE evaluate $M_i$
            \ENDFOR
        \ENDFOR
    \end{algorithmic}
\end{algorithm}

\subsection{Evaluating the Strong Lottery Ticket Hypothesis}\label{subsec:approach_strong}
We then evaluated the strong variant of the LTH by implementing a custom EA \cite{ncta24}. This EA leverages the same models as in the weak LTH experiments but bypasses the training step, focusing solely on learning a pruning mask. As before, we set a manual random seed, created an initial model with randomized weights, and based all subsequent models on these same weights. Adhering to the standard structure of an EA\cite{sloss2020ea_review}, our implementation follows the steps in \cref{alg:slth}: measure model performance, select top-performing models, and apply crossover, mutation, and migration to replenish the population. Although migration is typically used in parallel island setups\cite{harada2020parallel_ga}, we simulated it by introducing entirely new individuals. For the strong LTH, we measured both each individual's accuracy per generation and its number of remaining weights.

\begin{algorithm}
    \caption{Strong Lottery Ticket Hypothesis}
    \label{alg:slth}
    \begin{algorithmic}[1]
        \STATE \textbf{Input:} list of $random\_seeds$, number of generations $n\_gen$, number of individuals $n\_ind$
        
        \FOR{$random\_seed \in random\_seeds$}
            \STATE set $random\_seed$
            \STATE $M_0 \gets \text{generate initial model with empty pruning mask}$
            \STATE $individuals \gets$ create $n\_ind$ copies of $M_0$ and mutate their pruning masks
            
            \FOR{$i\_gen \in 1..n\_gen$}
                \STATE measure performances
                \STATE $individuals \gets$ select $33\%$
                \STATE $individuals \gets$ crossover (replenish $66\%$ of $n\_ind$)
                \STATE $individuals \gets$ mutation (replenish $95\%$ of $n\_ind$)
                \STATE $individuals \gets$ migration (replenish $100\%$ of $n\_ind$)
            \ENDFOR
            
            \STATE measure performances
        \ENDFOR
    \end{algorithmic}
\end{algorithm}

\subsection{Datasets}\label{sec:Datasets}
We selected the Iris dataset, introduced by Fisher in 1936\cite{Fisher1936Iris}, and the Wine dataset, introduced by Aeberhard et al. in 1994\cite{Aeberhard1992Wine}, to evaluate VQCs across different levels of complexity while keeping the datasets relatively small. The Iris dataset contains 150 instances of iris flowers with four features and three classes; we also derived a simplified version for binary classification by removing the third class. The Wine dataset has 178 instances, each with 13 properties divided into three classes, and we again created a simplified binary version by excluding the third class. These datasets, commonly used in classical ML and QML research, enabled us to explore the LTH on tasks ranging from relatively straightforward to moderately complex classification.

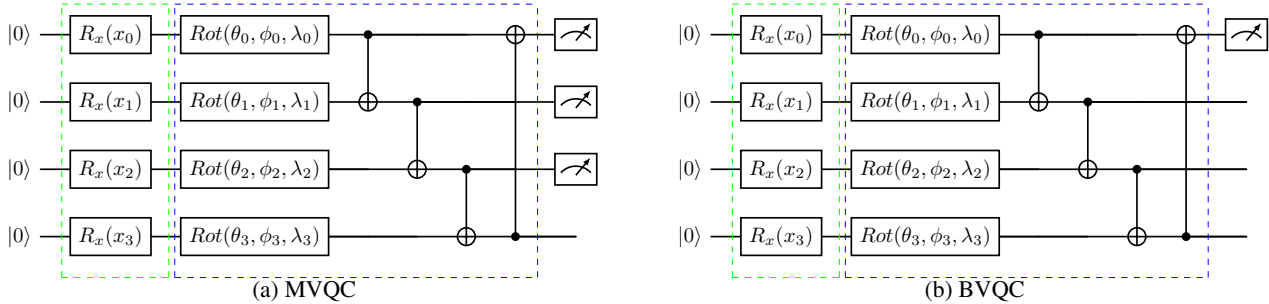
\begin{figure*}[htbp]
    \centering
    \begin{subfigure}[b]{0.48\linewidth}
        \centering
        \resizebox{\linewidth}{!}{ 
            \begin{quantikz}
                \lstick{\ket{0}} & \gate{R_x(x_0)} & \gate{Rot(\theta_0, \phi_0, \lambda_0)} & \ctrl{1} & \qw      & \qw       & \targ{}   & \meter{} \\
                \lstick{\ket{0}} & \gate{R_x(x_1)} & \gate{Rot(\theta_1, \phi_1, \lambda_1)} & \targ{}  & \ctrl{1} & \qw       & \qw       & \meter{} \\
                \lstick{\ket{0}} & \gate{R_x(x_2)} & \gate{Rot(\theta_2, \phi_2, \lambda_2)} & \qw      & \targ{}  & \ctrl{1}  & \qw       & \meter{} \\
                \lstick{\ket{0}} & \gate{R_x(x_3)} & \gate{Rot(\theta_3, \phi_3, \lambda_3)} & \qw      & \qw      & \targ{}   & \ctrl{-3} & \qw
            \end{quantikz}
            \begin{tikzpicture}[overlay, remember picture]
                \draw[dashed, green] (-9.2, -2.3) -- (-9.2, 2.3);
                \draw[dashed, green] (-9.2, 2.3) -- (-7.4, 2.3);
                \draw[dashed, green] (-7.4, -2.3) -- (-7.4, 2.3);
                \draw[dashed, green] (-9.2, -2.3) -- (-7.4, -2.3);
                
                \draw[dashed, blue] (-7.3, -2.3) -- (-7.3, 2.3);
                \draw[dashed, blue] (-7.3, 2.3) -- (-1.2, 2.3);
                \draw[dashed, blue] (-1.2, -2.3) -- (-1.2, 2.3);
                \draw[dashed, blue] (-7.3, -2.3) -- (-1.2, -2.3);
            \end{tikzpicture}
        }
        \caption{MVQC}
        \label{fig:mvqc}
    \end{subfigure}
    \hfill
    \begin{subfigure}[b]{0.48\linewidth}
        \centering
        \resizebox{\linewidth}{!}{ 
            \begin{quantikz}
                \lstick{\ket{0}} & \gate{R_x(x_0)} & \gate{Rot(\theta_0, \phi_0, \lambda_0)} & \ctrl{1} & \qw      & \qw       & \targ{}   & \meter{} \\
                \lstick{\ket{0}} & \gate{R_x(x_1)} & \gate{Rot(\theta_1, \phi_1, \lambda_1)} & \targ{}  & \ctrl{1} & \qw       & \qw       & \qw \\
                \lstick{\ket{0}} & \gate{R_x(x_2)} & \gate{Rot(\theta_2, \phi_2, \lambda_2)} & \qw      & \targ{}  & \ctrl{1}  & \qw       & \qw \\
                \lstick{\ket{0}} & \gate{R_x(x_3)} & \gate{Rot(\theta_3, \phi_3, \lambda_3)} & \qw      & \qw      & \targ{}   & \ctrl{-3} & \qw
            \end{quantikz}
            \begin{tikzpicture}[overlay, remember picture]
                \draw[dashed, green] (-9.2, -2.3) -- (-9.2, 2.3);
                \draw[dashed, green] (-9.2, 2.3) -- (-7.4, 2.3);
                \draw[dashed, green] (-7.4, -2.3) -- (-7.4, 2.3);
                \draw[dashed, green] (-9.2, -2.3) -- (-7.4, -2.3);
                \draw[dashed, blue] (-7.3, -2.3) -- (-7.3, 2.3);
                \draw[dashed, blue] (-7.3, 2.3) -- (-1.2, 2.3);
                \draw[dashed, blue] (-1.2, -2.3) -- (-1.2, 2.3);
                \draw[dashed, blue] (-7.3, -2.3) -- (-1.2, -2.3);
            \end{tikzpicture}
        }
        \caption{BVQC}
        \label{fig:bvqc}
    \end{subfigure}
    \caption{MVQC and BVQC, consisting of one wire per feature of the dataset. The green box shows angle embedding. The blue box symbolizes one layer, which is repeated several times. One wire per class of the dataset is measured.}
    \label{fig:vqcs}
\end{figure*}

\subsection{Models}\label{sec:Models}
We implemented three models to evaluate the LTH: two VQCs designed for binary and multi-class classification, and a classical NN baseline with a similar parameter count.

\paragraph{Multi-class VQC}
The primary model in this work is a multi-class VQC (MVQC) that uses angle encoding to map classical data into the quantum state space. The circuit includes strongly entangled layers to leverage quantum parallelism. We treated data re-uploading as a hyperparameter to reprocess inputs at multiple layers (\cref{subsec:hyperparameter}). The number of qubits was determined by the dataset’s number of features, using one wire per feature. The circuit measures the Pauli-Z expectation values of the first $n$ qubits (where $n$ is the number of classes) and applies a softmax function to convert these values into class probabilities. We trained the model with cross-entropy loss and the Adam optimizer. See \cref{fig:mvqc}.

\paragraph{Binary VQC}
For binary classification, we implemented a VQC (BVQC) that outputs the expectation value of a single qubit. We trained it using binary cross-entropy loss with logits and Adam optimization. See \cref{fig:bvqc}.

\paragraph{Simple Classical NN}
We also implemented a simple NN (SNN) as a baseline, consisting of a single hidden layer with 24 nodes and ReLU activation. The output layer applies softmax to produce class probabilities. We chose its size to match the VQCs’ parameter count (see \cref{tab:parameter_count}). It serves as a classical benchmark.

\subsection{Implementation}\label{sec:implementation}
In this section, we describe the Python-based implementation, discuss hyperparameter optimization, and outline the metrics used to evaluate results.

\subsubsection{Programming Language and Libraries}\label{subsec:prog_lang_and_libs}
All implementations were done in Python. For the ML framework, we used PyTorch\cite{Paszke2019PyTorch} for training, evaluation, and pruning. PennyLane\cite{Bergholm2018PennyLane} was employed for simulating quantum circuits and integrating them with PyTorch. We used scikit-learn\cite{Pedregosa2011Scikit} for data preprocessing and Seaborn\cite{Waskom2021Seaborn} for visualization. Hyperparameter optimization was conducted with Optuna\cite{Akiba2019Optuna}. We used seeds 0 through 9 to ensure reproducible outcomes and reduce the impact of random initialization.

\subsubsection{Hyperparameter Optimization}\label{subsec:hyperparameter}
We performed hyperparameter optimization with Optuna\cite{Akiba2019Optuna} to find an optimal configuration for the unpruned models, assuming a strong initial setup remains effective post-pruning. \cref{tab:hyperparameter_search} lists the hyperparameters and search ranges, while \cref{tab:hyperparameters} shows the final values selected for each model. The number of layers, data re-uploading flag, and uniform range apply only to VQCs. We manually adjusted the number of layers in some cases to ensure over-parameterization (\cref{tab:hyperparameters}).

\begin{center}
\begin{table}[htbp]
{
    \small
    \begin{center}
        \begin{tabular}[center]{lrr}
            \toprule
            Hyperparameter     & Min Value & Max Value \\
            \midrule
            Learning Rate      & 0.001     & 0.3 \\
            Weight Decay       & 0.0001    & 0.001 \\
            Number of Layers*  & 8         & 16 \\
            Data Re-Uploading  & False     & True \\
            Uniform Range      & $-\pi$    & $+\pi$ \\
            \bottomrule
        \end{tabular}
    \end{center}
}
\caption{Hyperparameters and their minimum/maximum values. *The \textit{number of layers} was manually adjusted in some models to ensure over-parameterization (\cref{tab:hyperparameters}).}
\label{tab:hyperparameter_search}
\end{table}
\end{center}

\subsection{Parameter Count}\label{sec:parameter_count}
The models differ in parameter counts due to architectural variations among SNN, BVQC, and MVQC. \cref{tab:parameter_count} lists the parameter counts for each model and dataset.

\begin{center}
\begin{table}[htbp]
{
    \small
    \begin{center}
        \begin{tabular}[center]{lrrrr}
            \toprule
            Dataset         & BVQC & MVQC & SNN \\
            \midrule
            Simplified Iris & 122  & 184  & 170 \\
            Iris            & -    & 198  & 195 \\
            Simplified Wine & 548  & 355  & 386 \\
            Wine            & -    & 630  & 411 \\
            \bottomrule
        \end{tabular}
    \end{center}
}
\caption{Number of parameters across models and datasets.}
\label{tab:parameter_count}
\end{table}
\end{center}

\subsection{Evaluation Metrics}\label{sec:evaluation_metrics}
We used accuracy on training and validation sets to gauge performance. For the weak LTH, we tracked accuracy at different levels of remaining weights. Each model/dataset combination has a curve at $100\%$ (unpruned) and at various pruned percentages. Minor rounding errors may arise due to integer parameter counts. For the strong LTH, we also recorded the percentage of remaining weights relative to accuracy per generation. The resulting plots for the weak LTH mirror those in the original study by Frankle and Carbin\cite{frankle2019lottery}, although the difference in problem size and architecture must be considered.

\section{Experiments}\label{sec:ResultsDiscussion}
In this Section, we present and interpret the findings of this work. For the weak LTH, we compare model performance under different pruning techniques and ratios. We use the multi-variate quantum circuit (MVQC) for multi-class tasks, the binary quantum circuit (BVQC) for binary tasks, and a standard neural network (SNN) with a parameter count comparable to the VQCs (see \cref{sec:Models}). The datasets include Iris and Wine, along with simplified versions reduced to two classes for binary classification (see \cref{sec:Datasets}). We employ iterative pruning and one-shot pruning (see \cref{subsec:approach_weak,subsec:approach_strong}) and compare performance both in the context of the weak LTH and strong LTH.

\subsection{Weak Lottery Ticket Hypothesis}\label{sec:result_weak}
For the weak LTH, we evaluate models across various pruning ratios, comparing fully-connected and pruned states. \cref{tab:winning_tickets} outlines the winning ticket pruning levels for different dataset-model combinations. Additional plots with curves down to 8\% remaining weights are available in \cref{sec:unselected_wlth}.

\begin{table}[htbp]
{
    \small
    \begin{center}
        \begin{tabular}[center]{lrrrr}
            \toprule
            Dataset         & BVQC   & MVQC   & SNN    \\
            \midrule
            Simplified Iris & 33.3\% & 26.1\% & 20.8\% \\
            Iris            & -      & 26.0\% & 32.1\% \\
            Simplified Wine & 51.3\% & /      & 51.4\% \\
            Wine            & -      & 32.7\% & 21.1\% \\
            \bottomrule
        \end{tabular}
    \end{center}
}
\caption{Percentage of remaining weights of winning tickets across models and datasets.}
\label{tab:winning_tickets}
\end{table}

\subsubsection{Comparison of Pruning Techniques}\label{subsec:comp_pruning_techniques}
\begin{figure}[htb]
 \centering
 \subfloat[][Iterative]{\includegraphics[width=0.5\linewidth]{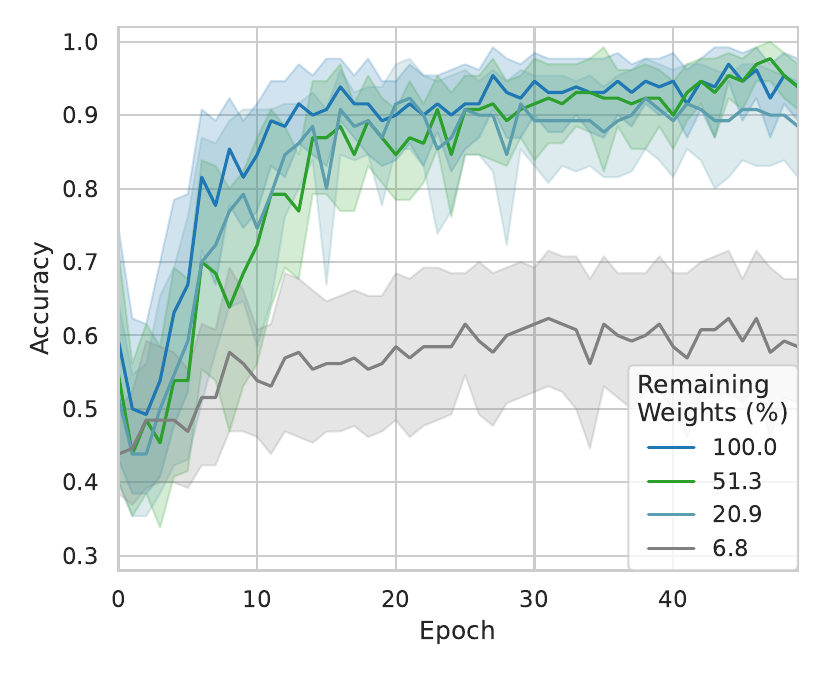}\label{fig:comparison_pruning_techniques_iterative}}
 \subfloat[][One-Shot]{\includegraphics[width=0.5\linewidth]{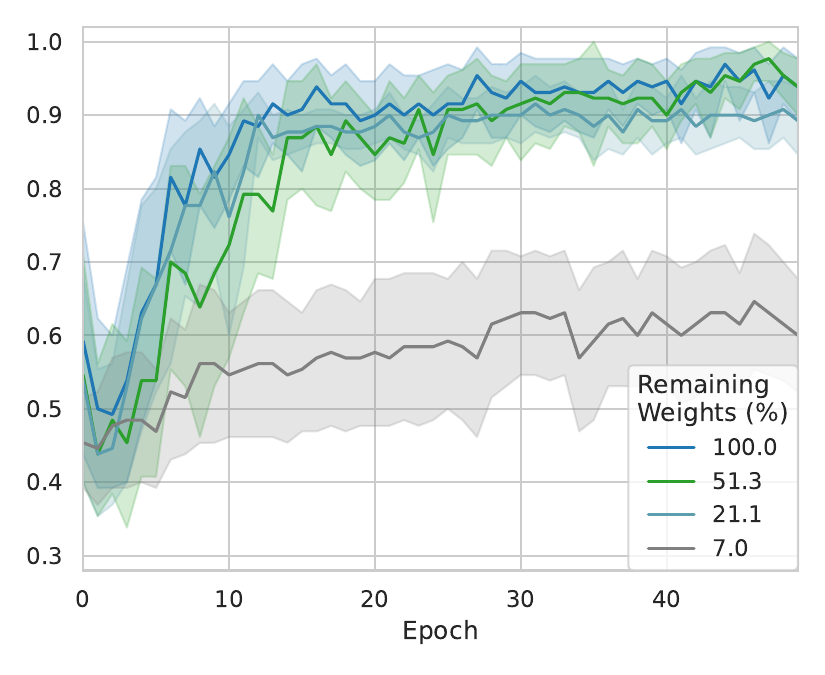}\label{fig:comparison_pruning_techniques_one_shot}}
 \caption{Comparison of pruning techniques on BVQC and simplified Wine.}
 \label{fig:comparison_pruning_techniques}
\end{figure}

As shown in \cref{fig:comparison_pruning_techniques}, there is no substantial difference between iterative and one-shot pruning. Small discrepancies likely stem from rounding differences in the exact number of remaining weights. Similar observations hold for all other dataset-model configurations and pruning ratios. We thus focus on iterative pruning in the subsequent results.

\subsubsection{Model Performance on Iris Dataset}\label{subsec:weak_iris}
\begin{figure}[htb]
 \centering
 \subfloat[][MVQC]{\includegraphics[width=0.5\linewidth]{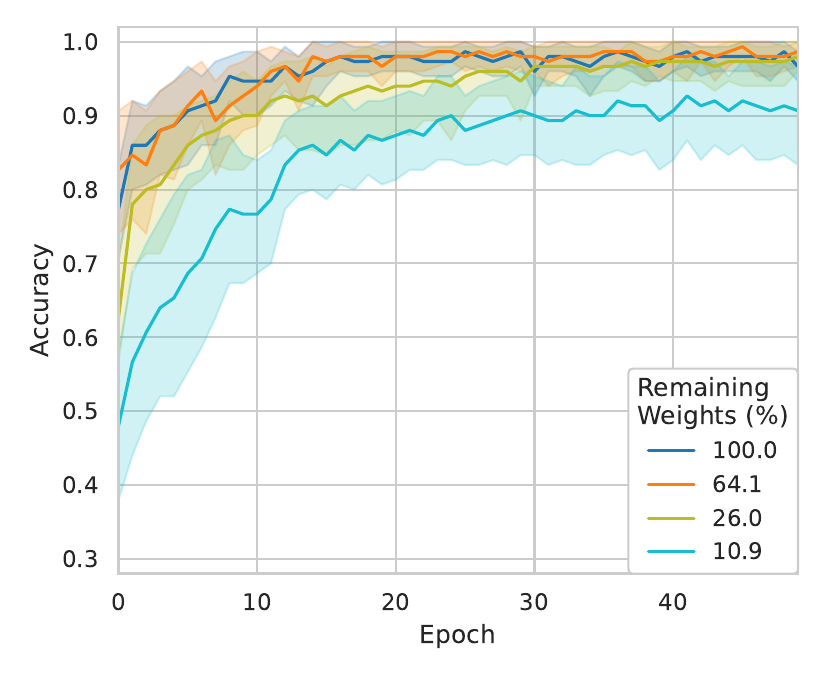}\label{fig:weak_3iris-mvqc}}
 \subfloat[][SNN]{\includegraphics[width=0.5\linewidth]{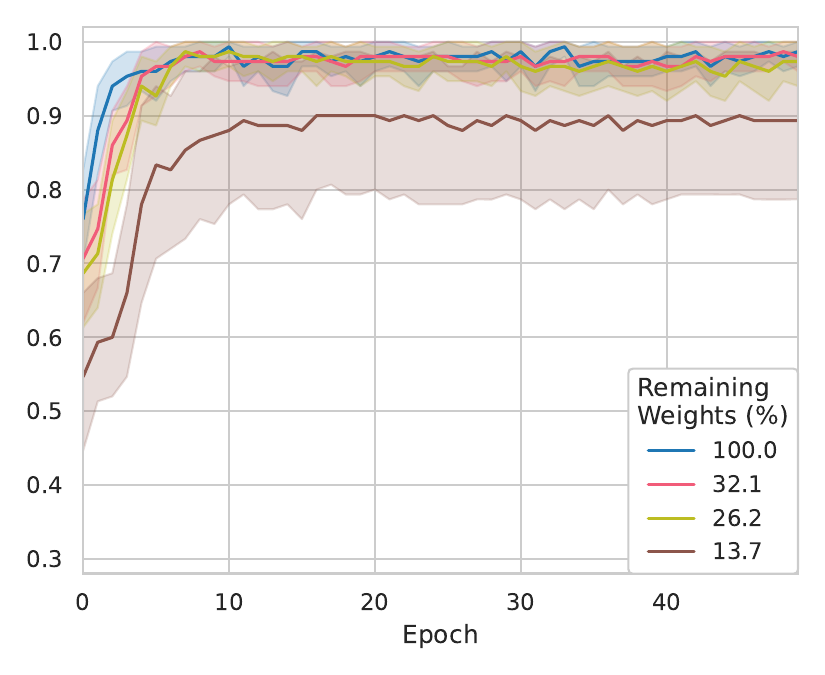}\label{fig:weak_3iris-snn}}
 \caption{Accuracies of MVQC and SNN on Iris.}
 \label{fig:weak_3iris}
\end{figure}

Since the Iris dataset has three classes, the BVQC is not applicable. \cref{fig:weak_3iris} shows that the MVQC and SNN each reach 97.5\% accuracy when unpruned. The MVQC keeps this accuracy down to 26.0\% remaining weights but falls below 90\% at 10.9\% remaining weights. The SNN similarly maintains 97.5\% accuracy at 32.1\% remaining weights, declining below 90\% at about 13.7\%.

\subsubsection{Model Performance on Simplified Iris Dataset}\label{subsec:weak_iris2}
\begin{figure}[htb]
 \centering
 \subfloat[][BVQC]{\includegraphics[width=0.33\linewidth]{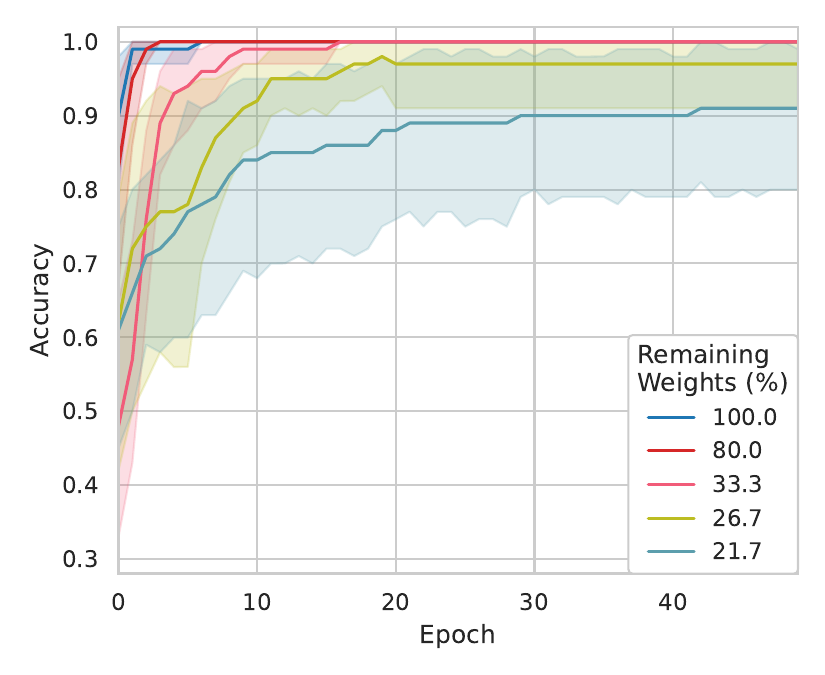}\label{fig:weak_2iris-bvqc}}
 \subfloat[][MVQC]{\includegraphics[width=0.33\linewidth]{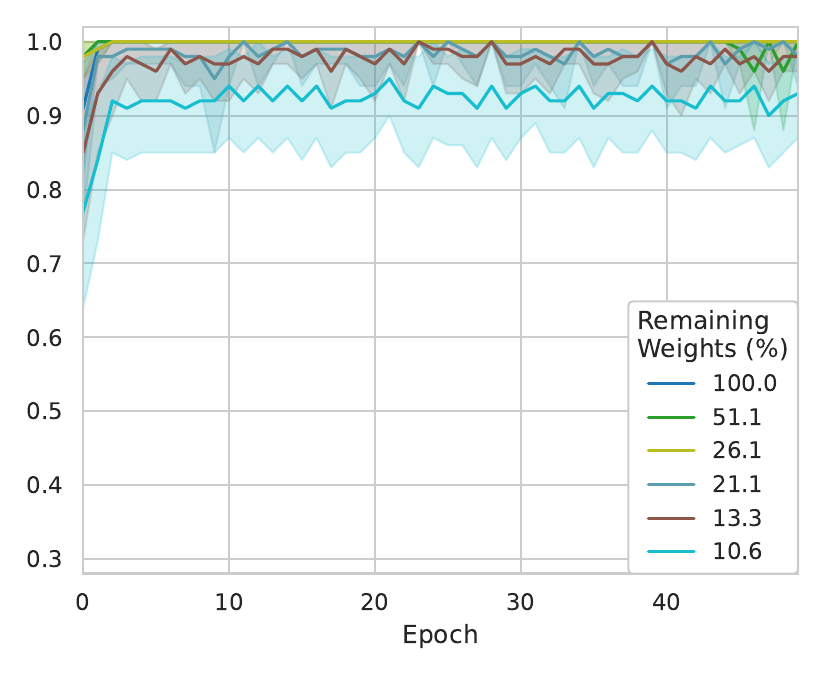}\label{fig:weak_2iris-mvqc}}
 \subfloat[][SNN]{\includegraphics[width=0.33\linewidth]{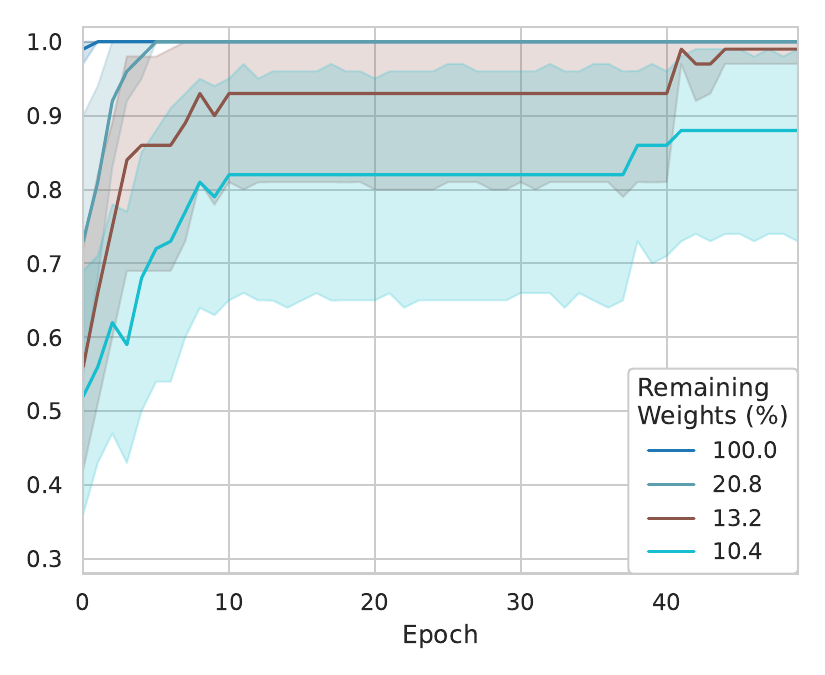}\label{fig:weak_2iris-snn}}
 \caption{Accuracies of BVQC, MVQC, and SNN on simplified Iris.}
 \label{fig:weak_2iris}
\end{figure}

\cref{fig:weak_2iris} presents results for the simplified Iris dataset (two classes). All three models reach 100\% accuracy in their unpruned states. The BVQC maintains 100\% accuracy down to 33.3\% remaining weights. The MVQC reaches 100\% down to 26.1\%, although some runs between 21.1\% and 13.3\% exhibit fluctuations. The SNN also retains 100\% accuracy until 20.8\% remaining weights.

\subsubsection{Model Performance on Wine Dataset}\label{subsec:wlth_wine}
\begin{figure}[htb]
 \centering
 \subfloat[][MVQC]{\includegraphics[width=0.5\linewidth]{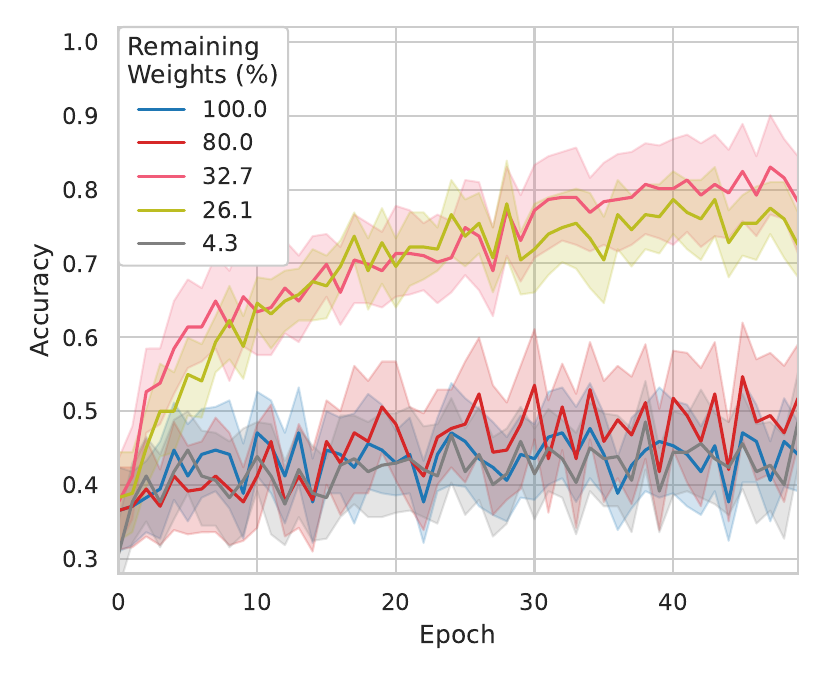}\label{fig:weak_3wine-mvqc}}
 \subfloat[][SNN]{\includegraphics[width=0.5\linewidth]{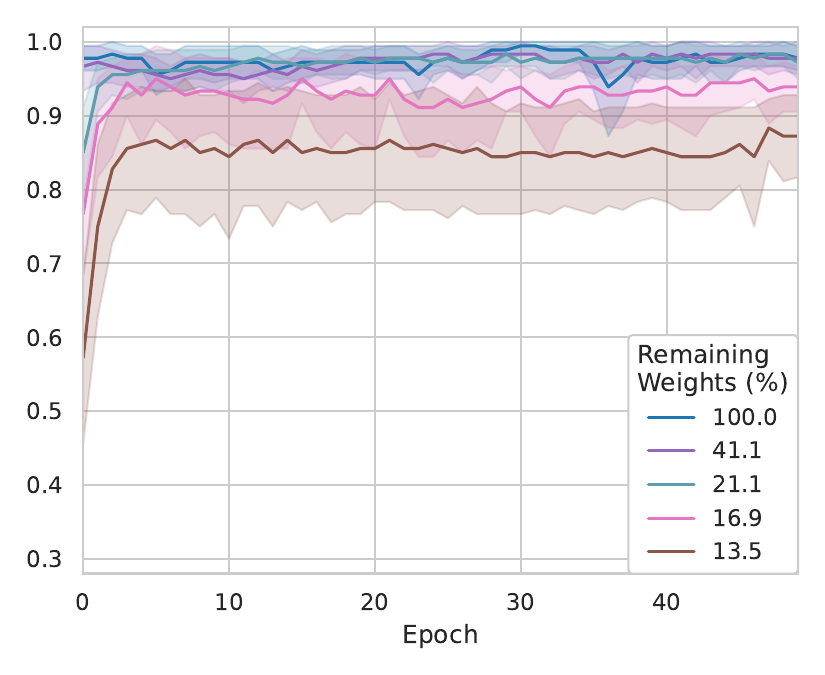}\label{fig:weak_3wine-snn}}
 \caption{Accuracies of MVQC and SNN on Wine.}
 \label{fig:weak_3wine}
\end{figure}

The Wine dataset contains three classes, so BVQC does not apply. \cref{fig:weak_3wine} shows the MVQC at 45\% accuracy when unpruned, increasing to 80\% at 32.7\% remaining weights. Even with only 4.3\% remaining weights, it surpasses its unpruned accuracy. The SNN starts at 98\% accuracy and remains near that level down to 21.1\%. Below about 16.9\% weights, its performance declines more steeply.

\subsubsection{Model Performance on Simplified Wine Dataset}\label{subsec:wlth_wine2}
\begin{figure}[htb]
 \centering
 \subfloat[][BVQC]{\includegraphics[width=0.33\linewidth]{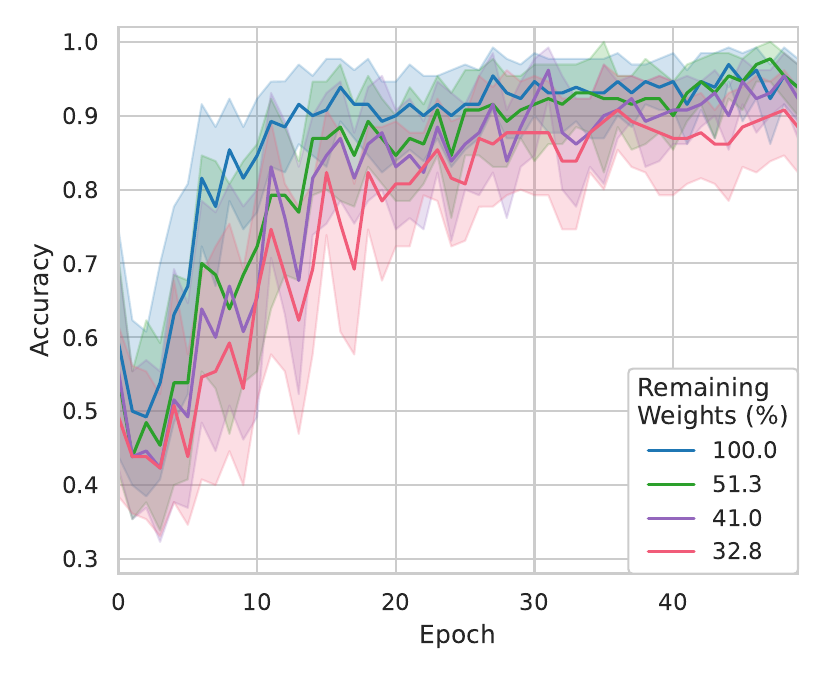}\label{fig:weak_2wine-bvqc}}
 \subfloat[][MVQC]{\includegraphics[width=0.33\linewidth]{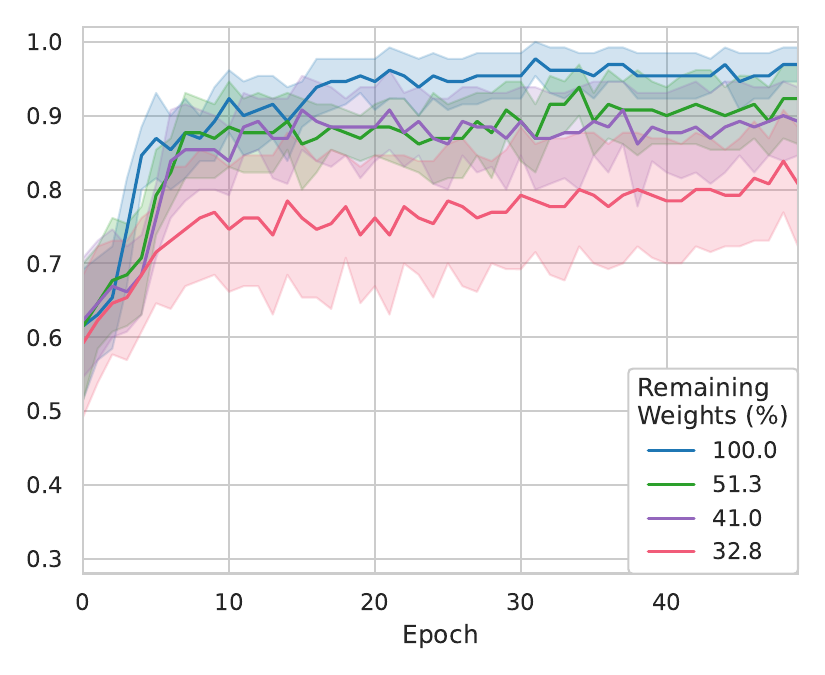}\label{fig:weak_2wine-mvqc}}
 \subfloat[][SNN]{\includegraphics[width=0.33\linewidth]{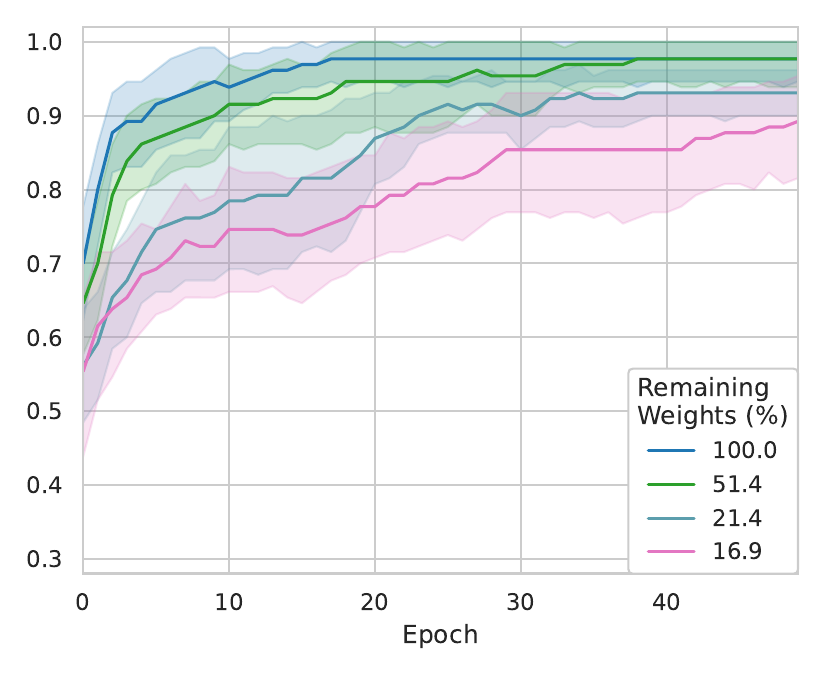}\label{fig:weak_2wine-snn}}
 \caption{Accuracies of BVQC, MVQC, and SNN on simplified Wine.}
 \label{fig:weak_2wine}
\end{figure}

\cref{fig:weak_2wine} illustrates that BVQC and MVQC both stabilize at about 94--96\% accuracy up to roughly 40\% remaining weights before declining, while SNN remains between 97--98\% accuracy until 51.4\% remaining weights. Pruning further causes noticeable performance drops, especially below about 20\% for the SNN.

\subsubsection{Discussion of Weak LTH}
Outcomes are consistent across iterative and one-shot pruning, possibly due to smaller problem sizes and simpler models than in prior research, such as Frankle and Carbin (2019) \cite{frankle2019lottery} or Liu et al. (2024) \cite{liu2024surveylotterytickethypothesis}. Despite this, both VQCs often retain high performance under substantial pruning. For example, the MVQC keeps winning tickets with about 26\% of the original weights on both the Iris and simplified Iris datasets. The BVQC and MVQC also maintain moderate performance when pruned for the Wine datasets, although the MVQC shows a more pronounced improvement after pruning on the unreduced Wine dataset, likely due to its highly over-parametrized nature. Across all datasets, the SNN shows strong performance and more stable training curves.

Although we do not see extremely low pruning levels (e.g., 3.6\%) reported by Frankle and Carbin (2019) \cite{frankle2019lottery}, these findings support the weak LTH for both classical and quantum networks.

\subsection{Strong Lottery Ticket Hypothesis}\label{sec:result_strong}
We next evaluate whether any pruned models surpass or match their unpruned counterparts (the strong LTH). Each dataset is assessed under an evolutionary algorithm (EA) approach. Plots compare average accuracies and remaining weights of the best individual per generation to weak LTH baselines at 100\% and a comparable remaining weight level.

\subsubsection{Model Performance on Iris Dataset}\label{subsec:strong_iris}
\begin{figure}[htb]
 \centering
 \subfloat[][MVQC]{\includegraphics[width=0.5\linewidth]{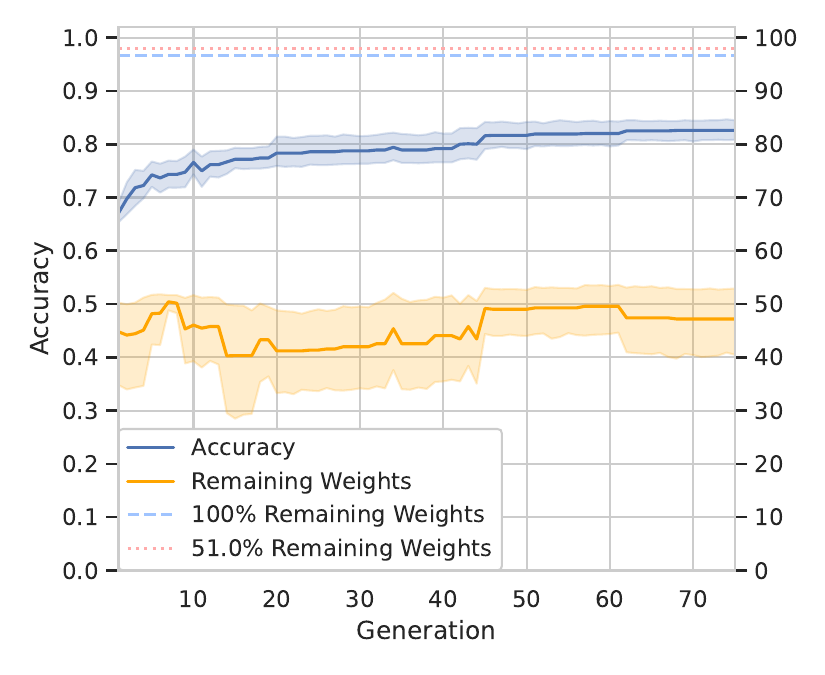}\label{fig:slth_3iris-mvqc}}
 \subfloat[][SNN]{\includegraphics[width=0.5\linewidth]{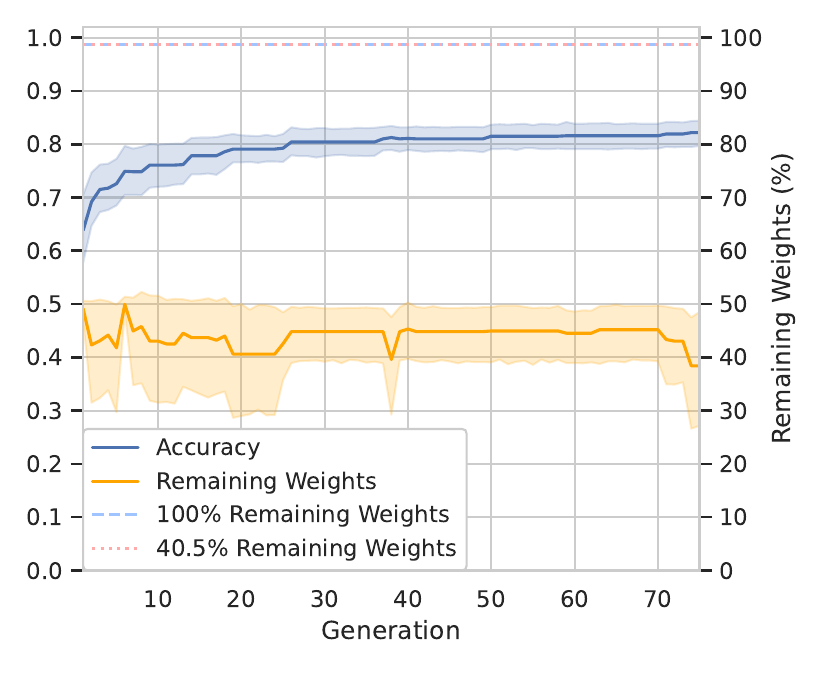}\label{fig:slth_3iris-snn}}
 \caption{Accuracies and remaining weights of best individual per generation of MVQC and SNN on Iris. Dashed lines show weak LTH results at specified remaining weights.}
 \label{fig:slth_3iris}
\end{figure}

As shown in \cref{fig:slth_3iris}, the MVQC improves from 68\% to 83\% through 20 generations, failing to reach the 97.5\% unpruned accuracy. The SNN similarly goes from 65\% to 82\%, also below 97.5\%. Both stabilize around 39--47\% remaining weights, yet models at similar pruning ratios under the weak LTH were more accurate.

\subsubsection{Model Performance on Simplified Iris Dataset}\label{subsec:strong_iris2}
\begin{figure}[htb]
 \centering
 \subfloat[][BVQC]{\includegraphics[width=0.33\linewidth]{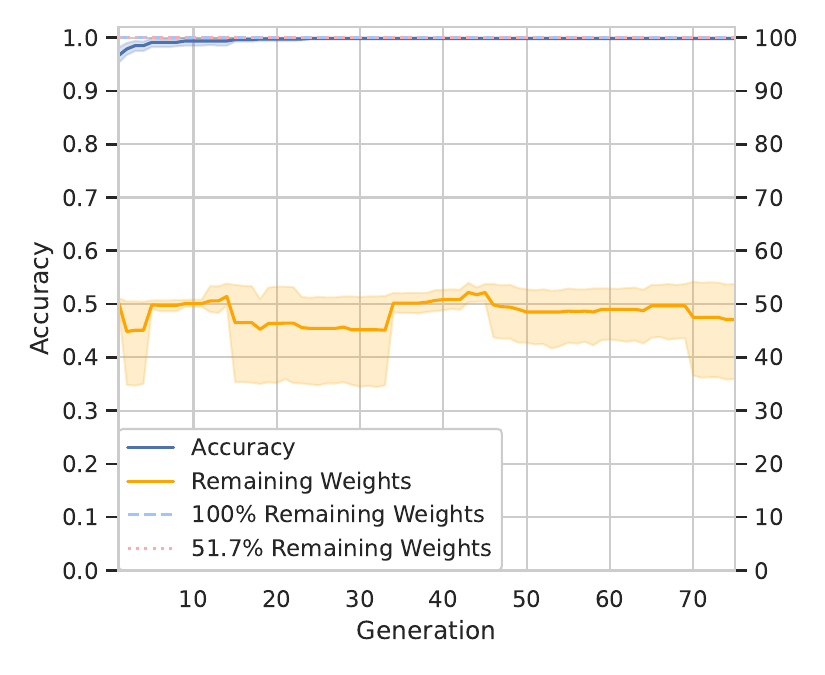}\label{fig:slth_2iris-bvqc}}
 \subfloat[][MVQC]{\includegraphics[width=0.33\linewidth]{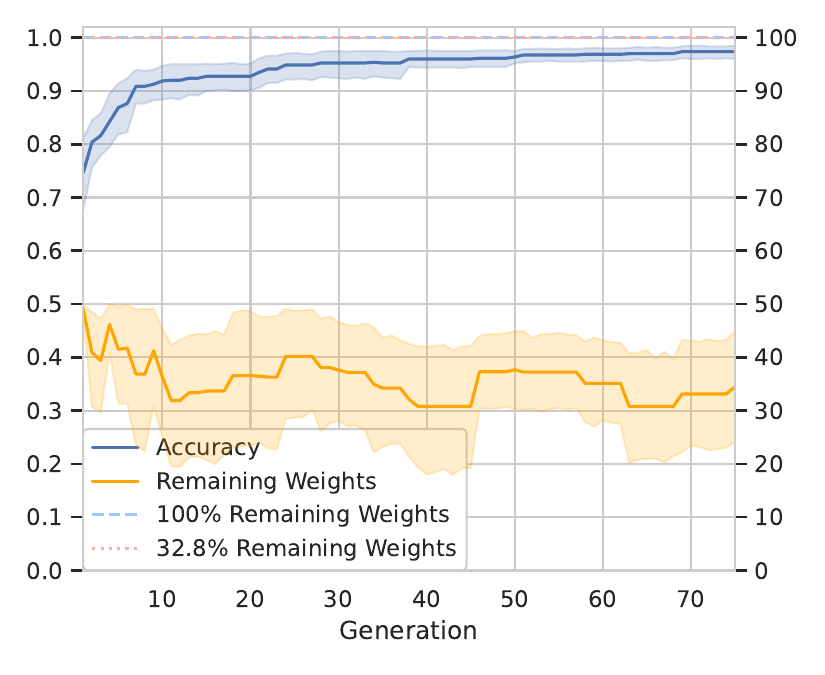}\label{fig:slth_2iris-mvqc}}
 \subfloat[][SNN]{\includegraphics[width=0.33\linewidth]{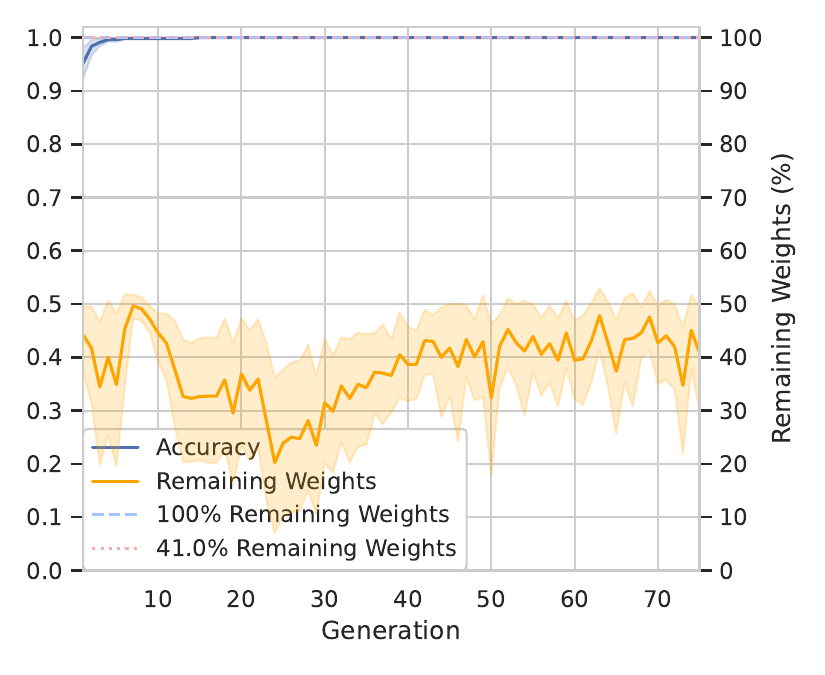}\label{fig:slth_2iris-snn}}
 \caption{Accuracies and remaining weights of best individual per generation on simplified Iris. Dashed lines show weak LTH results at specified remaining weights.}
 \label{fig:slth_2iris}
\end{figure}

In \cref{fig:slth_2iris}, the BVQC starts near 96\% and reaches 100\% after about 15 generations. Its final pruning ratio is about 48\%, comparable to a winning ticket under iterative pruning. The MVQC sees a notable gain from 75\% to 95\% accuracy at about 34\% remaining weights, but it still lags behind the unpruned MVQC’s 100\%. The SNN quickly reaches 100\% with about 40\% of the original weights, matching its unpruned accuracy.

\subsubsection{Model Performance on Wine Dataset}\label{subsec:strong_wine}
\begin{figure}[htbp]
 \centering
 \subfloat[][MVQC]{\includegraphics[width=0.5\linewidth]{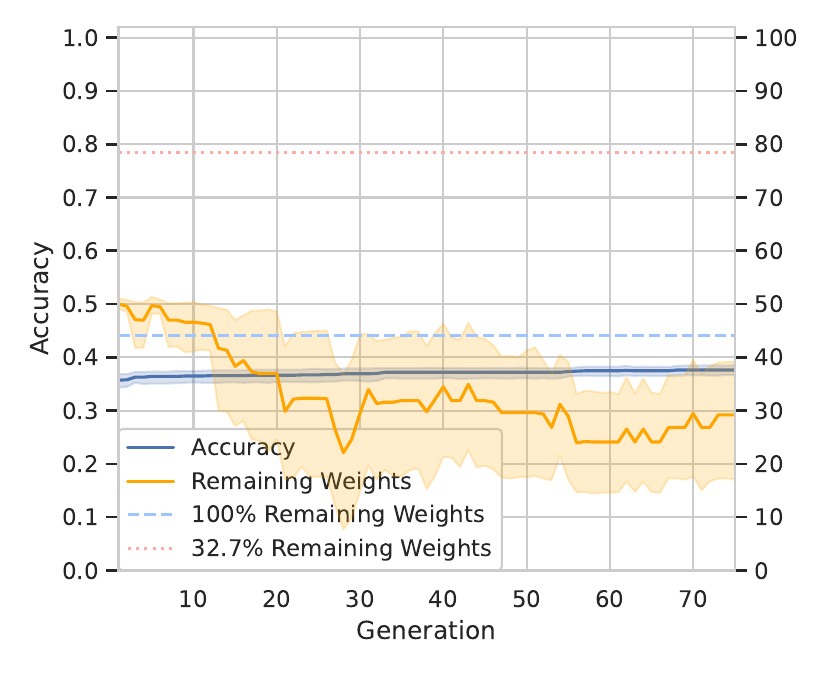}\label{fig:slth_3wine-mvqc}}
 \subfloat[][SNN]{\includegraphics[width=0.5\linewidth]{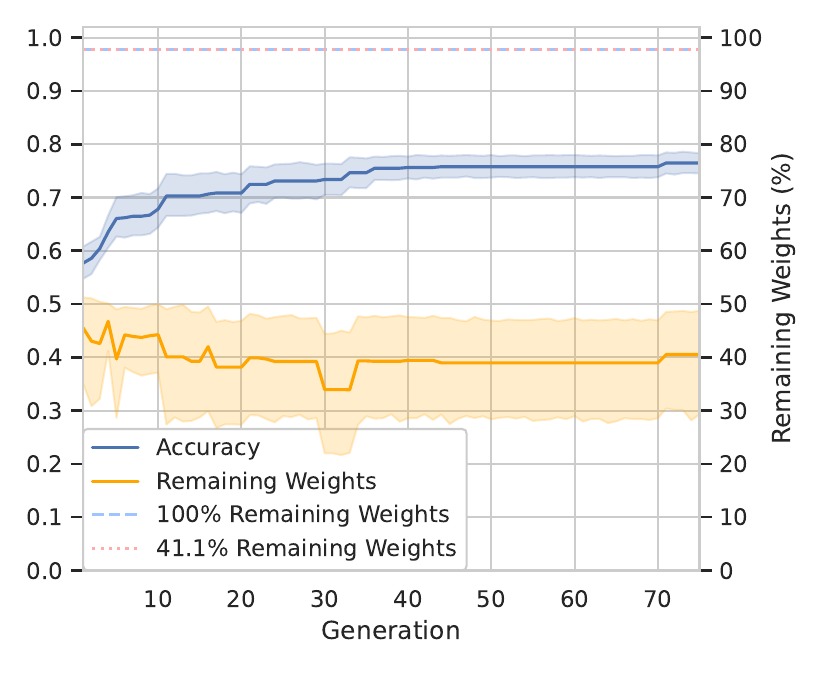}\label{fig:slth_3wine-snn}}
 \caption{Accuracies and remaining weights of best individual per generation of MVQC and SNN on Wine. Dashed lines show weak LTH results at specified remaining weights.}
 \label{fig:slth_3wine}
\end{figure}

\cref{fig:slth_3wine} shows that the MVQC reaches about 48\% accuracy and the SNN about 87\%, both falling short of their unpruned versions under the weak LTH (45\% vs.\ 80\% for MVQC; 87\% vs.\ 98\% for SNN). The remaining weights hover around 30--40\%, but this ratio does not reliably indicate the EA's success, as subnetwork size is not a direct proxy for accuracy. Further exploration is needed to better understand the relationship between pruning, subnetwork size, and model performance.

\subsubsection{Model Performance on Simplified Wine Dataset}\label{subsec:strong_wine2}
\begin{figure}[hbt]
 \centering
 \subfloat[][BVQC]{\includegraphics[width=0.33\linewidth]{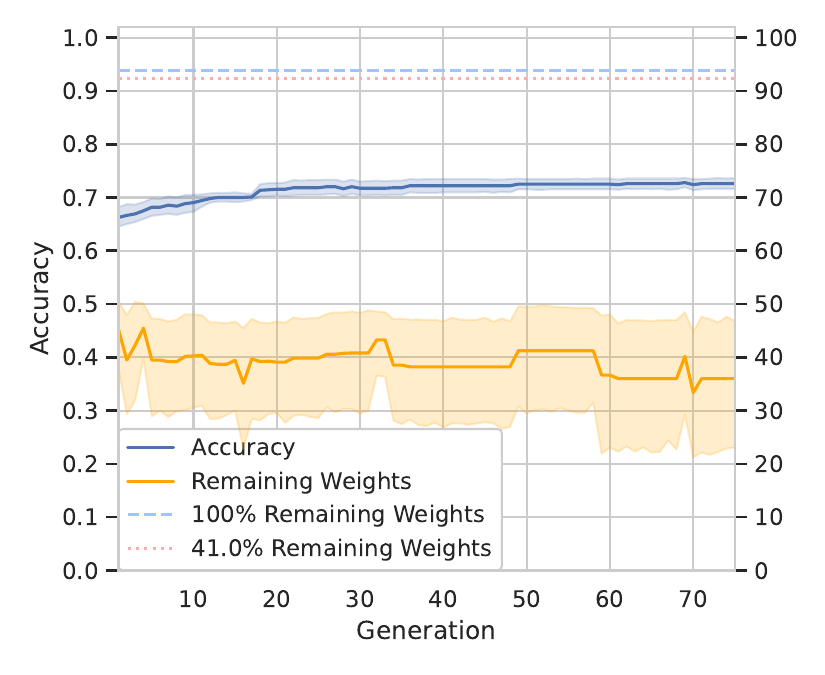}\label{fig:slth_2wine-bvqc}}
 \subfloat[][MVQC]{\includegraphics[width=0.33\linewidth]{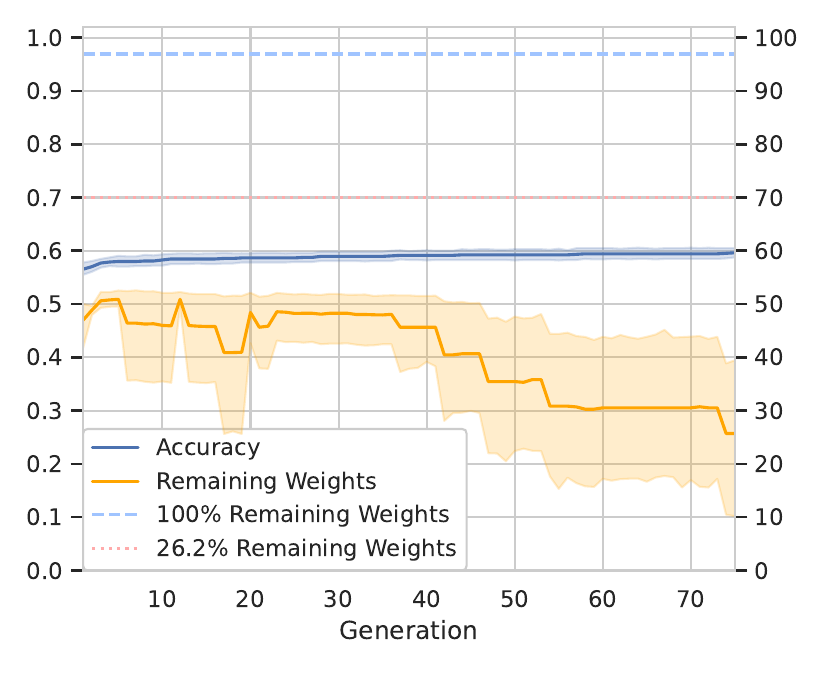}\label{fig:slth_2wine-mvqc}}
 \subfloat[][SNN]{\includegraphics[width=0.33\linewidth]{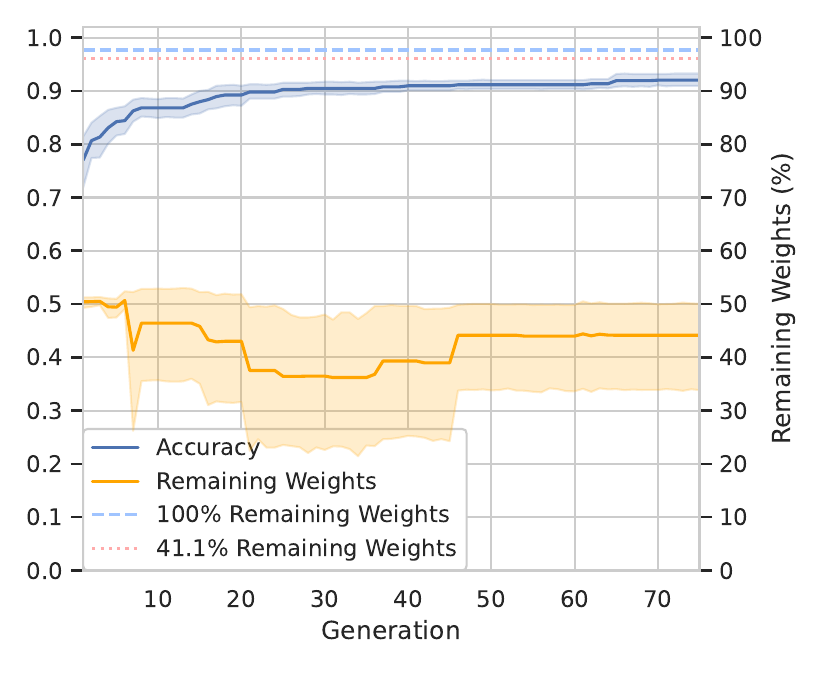}\label{fig:slth_2wine-snn}}
 \caption{Accuracies and remaining weights of best individual per generation on simplified Wine. Dashed lines show weak LTH results at specified remaining weights.}
 \label{fig:slth_2wine}
\end{figure}

\cref{fig:slth_2wine} reveals that the BVQC, MVQC, and SNN remain well below their unpruned accuracies (94\% BVQC, 97\% MVQC, and 98\% SNN under the weak LTH). The best EA solutions stabilize around 44\% weights for the SNN and 26--37\% for the quantum models, indicating the EA could not identify subnetworks that match or exceed the full models.

\subsubsection{Discussion of Strong LTH}
On the simplified Iris dataset, the EA discovers winning tickets that match unpruned accuracies for the BVQC and SNN; the MVQC nearly achieves its unpruned performance. This suggests the EA can find effective masks for VQCs in simpler tasks. However, for the more complex or higher-dimensional Wine dataset, no winning tickets emerge. The EA yields subnetworks with 30--45\% of the weights but fails to reach unpruned accuracy. Different optimization techniques or more flexible evolutionary strategies might uncover stronger subnetworks for more challenging tasks.

\section{Conclusion}\label{sec:Conclusion}
In this work, we applied the LTH to VQCs to investigate whether winning tickets can emerge, similar to those in classical NNs. We employed iterative and one-shot pruning to evaluate the weak LTH, as well as an EA to assess the strong LTH. We used two datasets, Iris and Wine, both with simplified variants for binary classification. Additionally, we tested two VQCs—one designed for multi-class classification and another for binary classification—and a simple NN as a baseline.

The experiments show that both the BVQC and MVQC achieved winning tickets with as few as 33.3\% and 26.0\% of remaining weights, respectively, using iterative pruning on the Iris and simplified Iris datasets. On these datasets, the MVQC performed comparably to an NN with a similar parameter count. On larger problem sizes, such as the simplified Wine dataset, both models had difficulty finding winning tickets; however, the BVQC still discovered one with 51.3\% remaining weights. In contrast, no winning ticket was found for the MVQC on the simplified Wine dataset, whereas the SNN, a relatively small NN with a parameter count comparable to that of the MVQC, still exhibited winning tickets. Nonetheless, on the unreduced Wine dataset, the MVQC illustrated that the LTH can overcome BPs in VQCs, yielding a 35\% performance improvement and reaching 80\% accuracy when pruned to 32.7\% of the original weights, whereas the unpruned model achieved only 45\%.

These findings indicate that the weak LTH is applicable to VQCs, and that the identified winning tickets can compete with those in NNs of similar size if the problem size is small enough. They also suggest that BPs can be mitigated if the initially unpruned model is sufficiently over-parameterized. Interestingly, we observed that iterative and one-shot pruning methods identified the same subnetworks and subcircuits, potentially due to the relatively small size and complexity of both the datasets and the models. This factor makes it challenging to compare these results with other studies, which generally involve larger datasets and models.

Regarding the strong LTH, we found a winning ticket for the BVQC on the simplified Iris dataset, retaining 100\% accuracy with 45\% of the weights remaining. We also observed a substantial 20\% improvement—from 75\% to 95\% accuracy—for the MVQC on the simplified Iris dataset with 40\% remaining weights. Although this does not match the MVQC’s 100\% accuracy at full capacity, it remains a noteworthy result.

Future work could investigate the LTH on real quantum hardware and extend it to larger circuits and problem sizes. It might also explore pruning gates rather than just weights, following concepts like Structure Symmetric Pruning\cite{ma2024continuous}. As part of addressing BPs, further studies could focus on scenarios in which BPs are more pronounced and examine how strongly pruned VQCs might overcome them.




\section*{Impact Statement}

This paper advances our understanding of how the LTH may enhance efficiency in quantum machine learning. By showing that VQCs can be pruned to sparse “winning tickets” without compromising performance, our work highlights a pathway to reduce computational and resource overhead in emerging quantum technologies. A more resource-efficient approach to VQCs could lower energy consumption and hardware demands, fostering broader accessibility to quantum methods. Additionally, it may accelerate real-world applications of quantum machine learning in fields ranging from drug discovery to secure communications. We do not identify direct ethical or societal risks at this stage, but continued research should remain attentive to potential long-term impacts, such as shifts in computational power and its implications for data security and privacy.

\bibliography{main}
\bibliographystyle{icml2025}

\newpage
\appendix
\onecolumn
\section{Hyperparameter Search Results}\label{sec:hyperparameter_search_results}
\cref{tab:hyperparameters} shows the results of the hyperparameter search (see \cref{subsec:hyperparameter}) and thus the final hyperparameter values for the models used in the experiments. The hyperparameters \textit{learning rate} and \textit{weight decay} have been used for all three models, while \textit{number of layer}, \textit{data re-uploading} and \textit{uniform range} have been used for the VQCs only. The \textit{number of layers} have been manually adjusted for the MVQC on the unreduced Iris and Wine datasets and for the BVQC on the simplified Iris dataset to ensure over-parametrization.

\renewcommand{\arraystretch}{1.2}

    \begin{table}[htbp]
    \centering
    {
        \begin{tabular}{|r|l|r|r|r|r|r|l|}
            \thickhline
            \textbf{Dataset} & \textbf{Model} & \textbf{Learning Rate} & \textbf{Weight Decay} & \textbf{L}\tablefootnote{L = Number of Layers (only for VQCs)} & \textbf{DRU}\tablefootnote{DRU = Data Re-Uploading (only for VQCs)} & \textbf{Uniform Range}\tablefootnote{only for VQCs}\\
            \hline
            S. Iris & BVQC & 0.0061690543775444456 & 0.00011451748647630793 & 10* & False & 1.78340734641020670 \\
            S. Iris & MVQC & 0.1404603283295513300 & 0.00020043419481312650 & 15  & False & 0.34099999999999997 \\
            S. Iris & SNN  & 0.0189753941329335250 & 0.00037958686849631810 & -   & -     & - \\
            \hline
            Iris & MVQC    & 0.0276674656133278770 & 0.00014188599748059832 & 16* & False & 0.99700000000000000 \\
            Iris & SNN     & 0.0412876064499254560 & 0.00011458294311477400 & -   & -     & - \\
            \hline
            S. Wine & BVQC & 0.0341163513021525840 & 0.00048832766322303640 & 14  & False & 0.76659265358979310 \\
            S. Wine & MVQC & 0.0469360377064666600 & 0.00017400041959447874 & 9   & False & 0.13840734641020713 \\
            S. Wine & SNN  & 0.0012576386169755418 & 0.00077375502517089950 & -   & -     & - \\
            \hline
            Wine & MVQC    & 0.0562921735356738800 & 0.00033968499286871637 & 16* & False & 0.35300000000000000 \\
            Wine & SNN     & 0.0574452822141239800 & 0.00010743993876395757 & -   & -     & - \\
            \thickhline
        \end{tabular}
        }
        \caption{Hyperparameters values across models and datasets. *The \textit{number of layers} have been manually adjusted for some models to ensure over-parametrization.}
        \label{tab:hyperparameters}
    \end{table}

\renewcommand{\arraystretch}{1}

\section{Weak Lottery Ticket Hypothesis - Full Plots}\label{sec:unselected_wlth}
The following figures show the results of the weak LTH utilizing iterative pruning across all models and dataset in an unselected version, with the curves for all levels of remaining weights above 8\%.

\begin{figure}[htb]
 \centering
 \subfloat[][MVQC]{\includegraphics[width=0.5\textwidth]{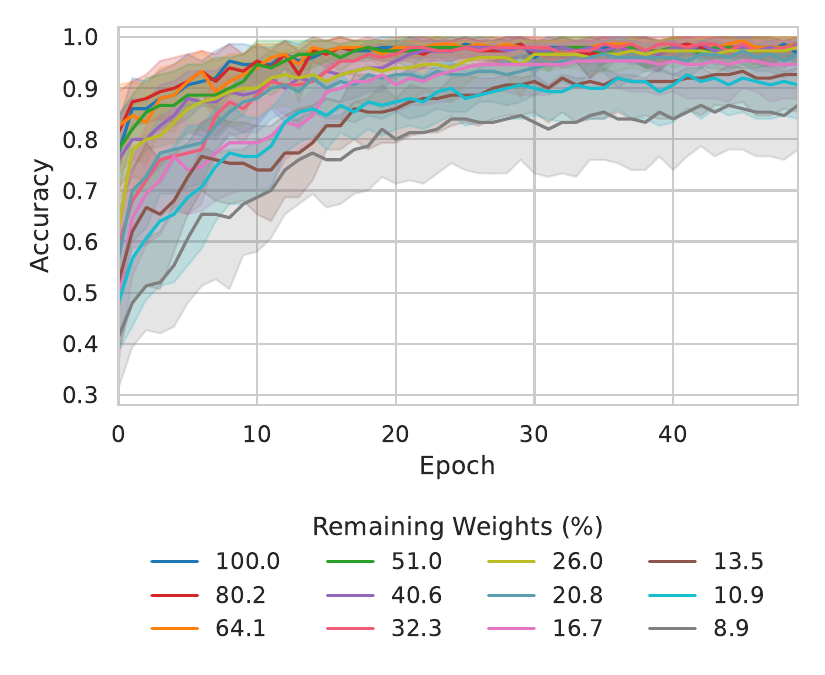}\label{fig:weak_3iris-mvqc_unselected}}
 \subfloat[][SNN]{\includegraphics[width=0.5\textwidth]{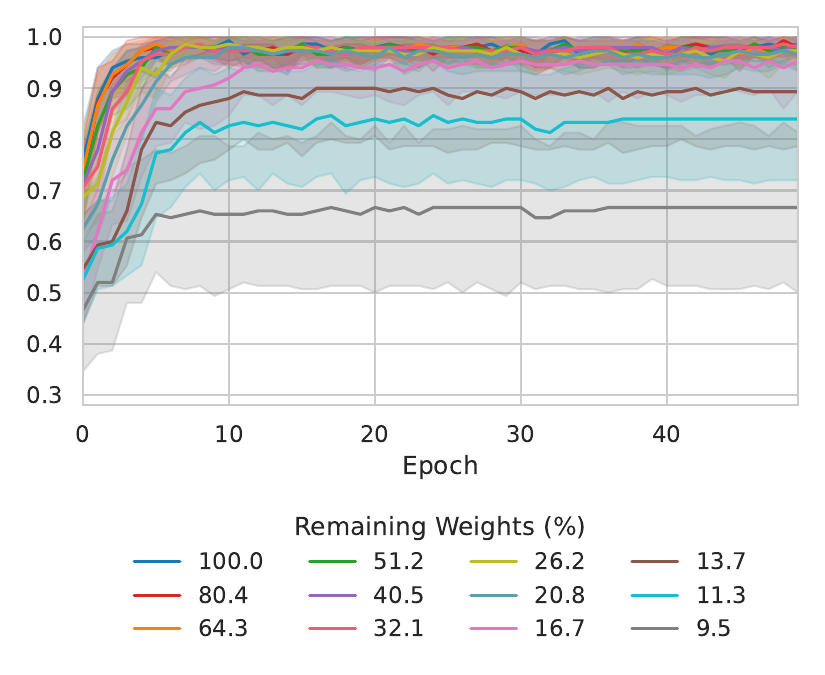}\label{fig:weak_3iris-snn_unselected}}
 \caption{Accuracies of MVQC \& SNN on Iris, all levels of remaining weights down to 8\%.}
 \label{fig:weak_3iris_unselected}
\end{figure}

\begin{figure}[htb]
 \centering
 \subfloat[][BVQC]{\includegraphics[width=0.33\textwidth]{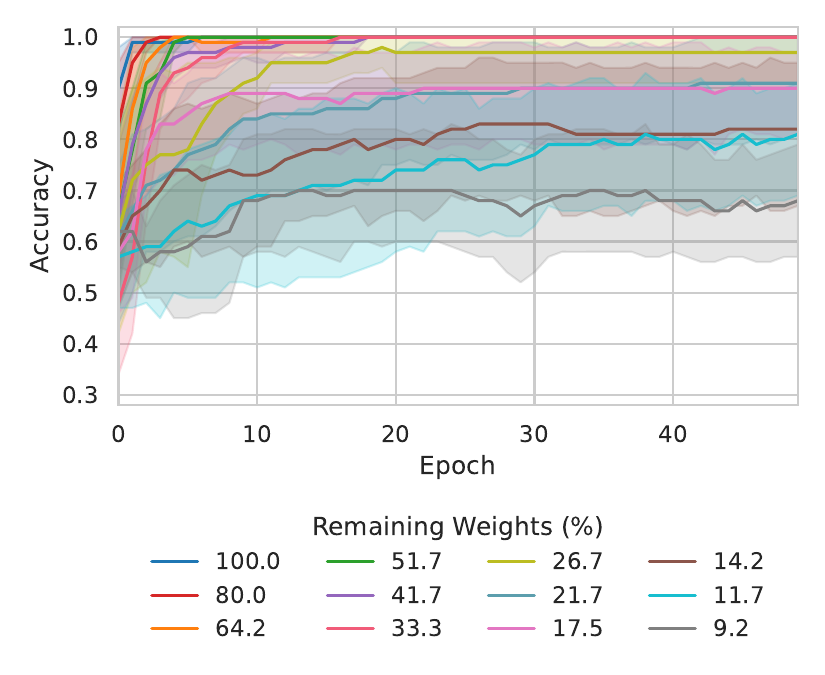}\label{fig:weak_2iris-bvqc_unselected}}
 \subfloat[][MVQC]{\includegraphics[width=0.33\textwidth]{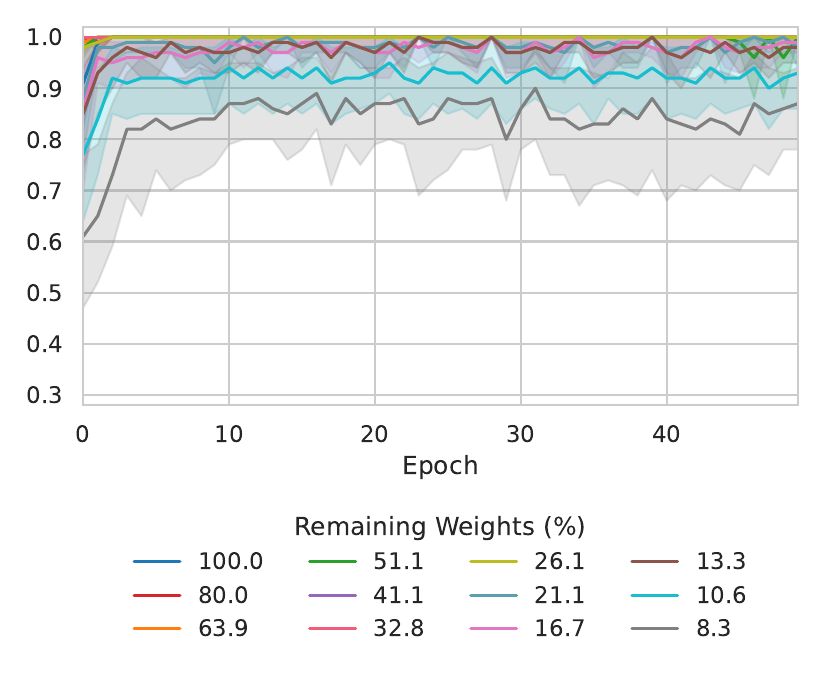}\label{fig:weak_2iris-mvqc_unselected}}
 \subfloat[][SNN]{\includegraphics[width=0.33\textwidth]{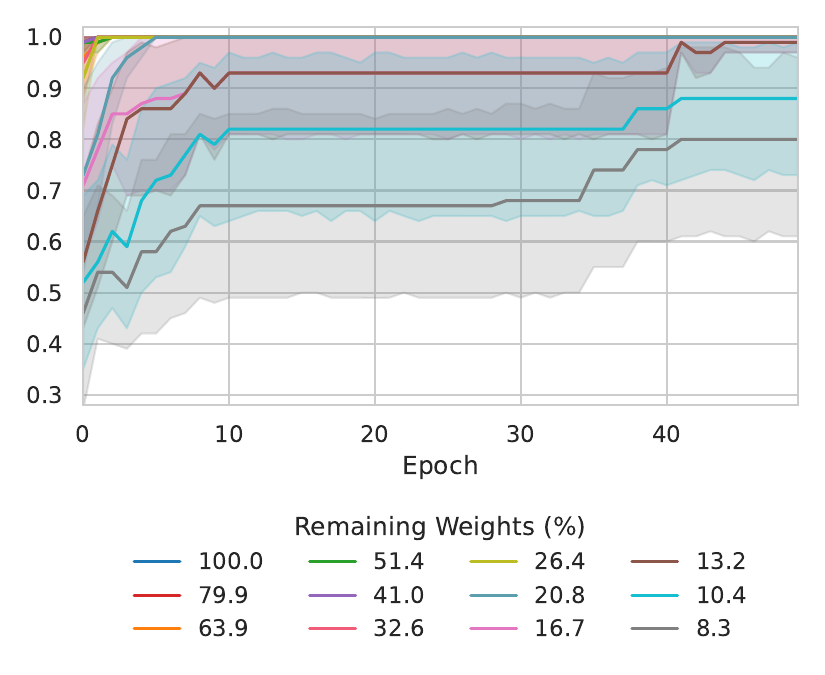}\label{fig:weak_2iris-snn_unselected}}
 \caption{Accuracies of BVQC, MVQC \& SNN on simplified Iris, all levels of remaining weights down to 8\%.}
 \label{fig:weak_2iris_unselected}
\end{figure}

\begin{figure}[htb]
 \centering
 \subfloat[][MVQC]{\includegraphics[width=0.5\textwidth]{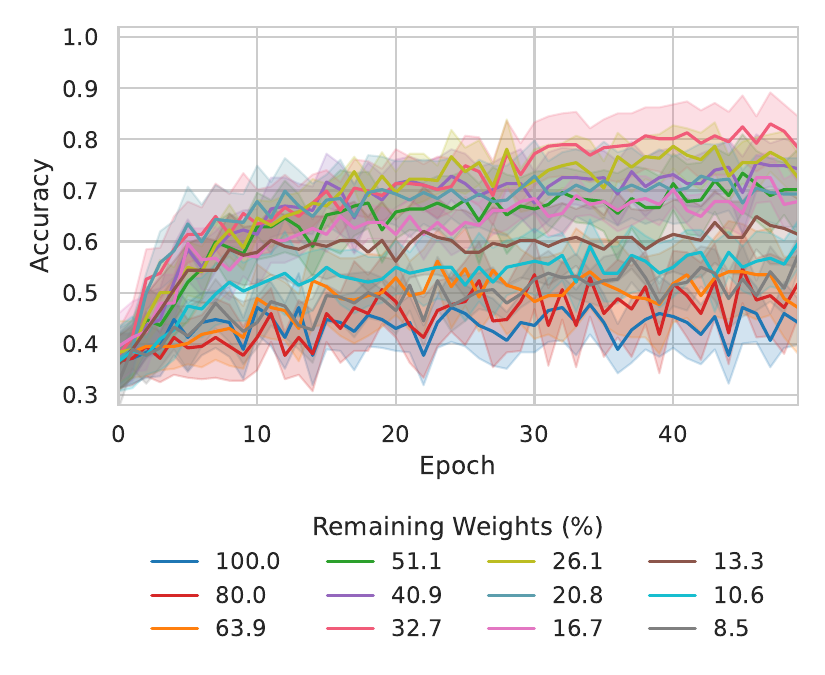}\label{fig:weak_3wine-mvqc_unselected}}
 \subfloat[][SNN]{\includegraphics[width=0.5\textwidth]{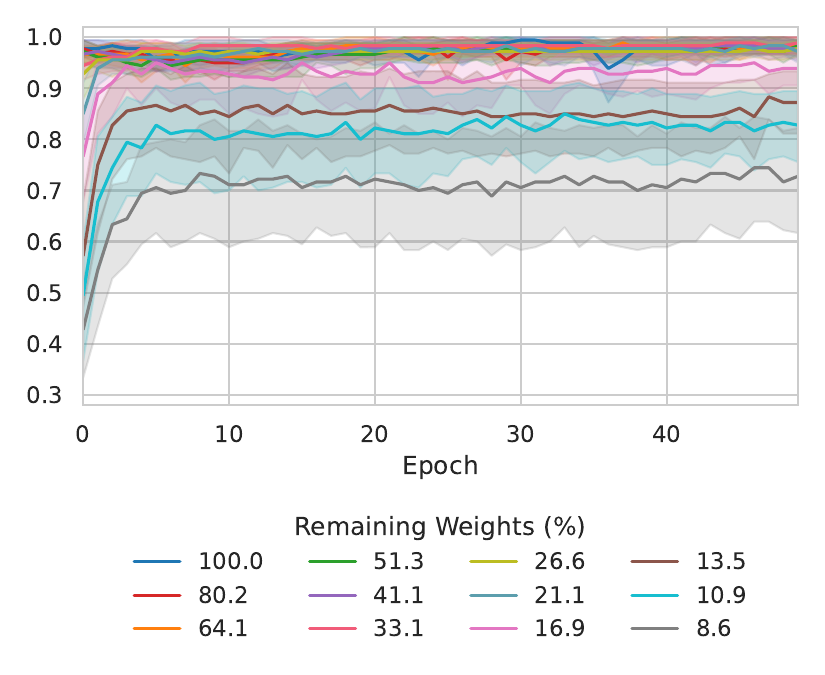}\label{fig:weak_3wine-snn_unselected}}
 \caption{Accuracies of MVQC \& SNN on Wine, all levels of remaining weights down to 8\%.}
 \label{fig:weak_3wine_unselected}
\end{figure}

\begin{figure}[htb]
 \centering
 \subfloat[][BVQC]{\includegraphics[width=0.33\textwidth]{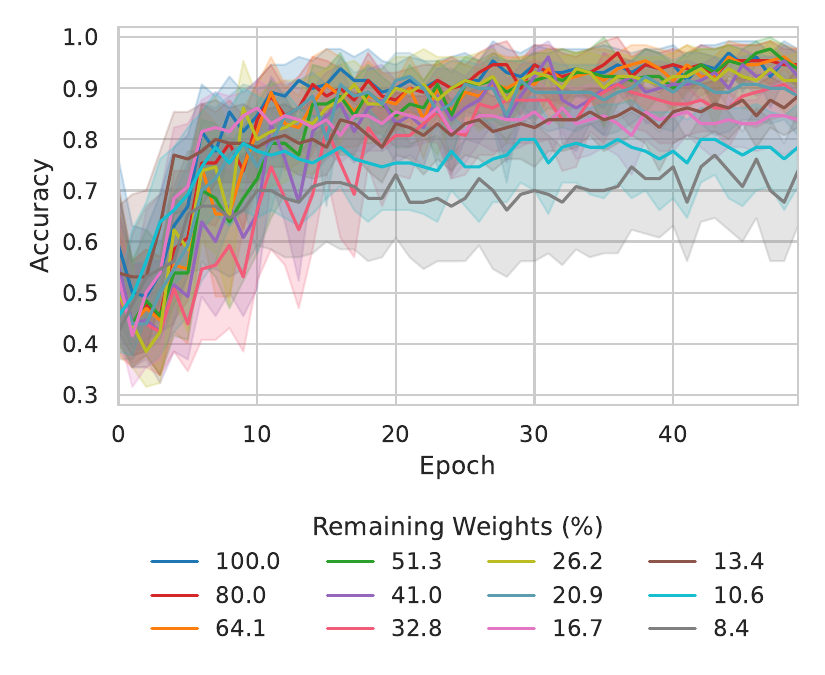}\label{fig:weak_2wine-bvqc_unselected}}
 \subfloat[][MVQC]{\includegraphics[width=0.33\textwidth]{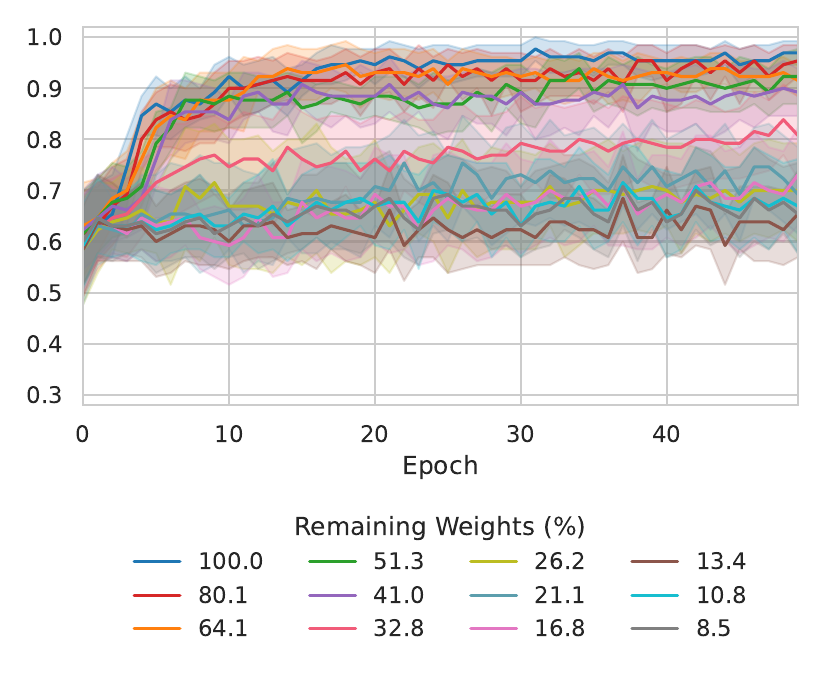}\label{fig:weak_2wine-mvqc_unselected}}
 \subfloat[][SNN]{\includegraphics[width=0.33\textwidth]{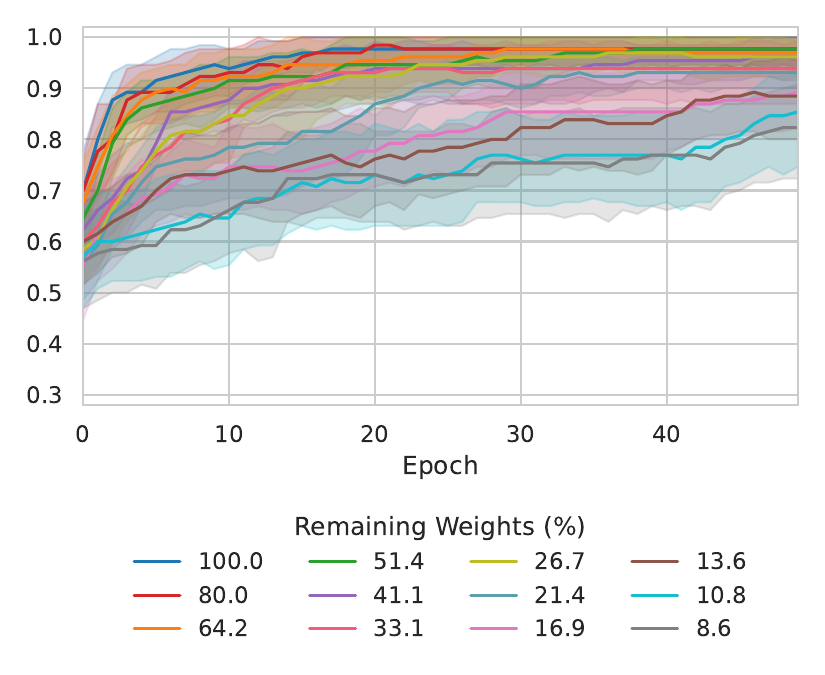}\label{fig:weak_2wine-snn_unselected}}
 \caption{Accuracies of BVQC, MVQC \& SNN on simplified Wine, all levels of remaining weights down to 8\%.}
 \label{fig:weak_2wine_unselected}
\end{figure}


\end{document}